\documentclass[manuscript,screen]{acmart}

\AtBeginDocument{%
  \providecommand\BibTeX{{%
    \normalfont B\kern-0.5em{\scshape i\kern-0.25em b}\kern-0.8em\TeX}}}

\usepackage{amsmath,amsfonts}
\usepackage{algorithmic}
\usepackage{graphicx}
\usepackage{textcomp}
\usepackage{xcolor}

\usepackage{subfigure}
\usepackage{multirow}
\usepackage{soul}

\begin{document}

\title{NAS-Cap: Deep-Learning Driven 3-D Capacitance Extraction with Neural Architecture Search and Data Augmentation
}

\thanks{The preliminary version has been presented at the Proc. Design, Automation \& Test in Europe Conference (DATE) in 2024. 
H. Li and D. Yang contributed equally to this work.}
\def\Tsinghua{
\affiliation{
  \institution{Department of Computer Science and Technology, BNRist, Tsinghua University}
  \city{Beijing}
  \country{China}
  \postcode{100084}
}
}

\author{Haoyuan Li}
\email{lihaoyua17@mails.tsinghua.edu.cn}
\author{Dingcheng Yang}
\email{ydc19@mails.tsinghua.edu.cn}
\author{Chunyan Pei}
\email{peicy@tsinghua.edu.cn}
\author{Wenjian Yu}
\email{yu-wj@tsinghua.edu.cn}
\Tsinghua


\begin{abstract}
More accurate capacitance extraction is demanded for designing integrated circuits under advanced process technology. The pattern matching approach and the field solver for capacitance extraction have the drawbacks of inaccuracy and large computational cost, respectively. Recent work \cite{yang2023cnn} proposes a grid-based data representation and a convolutional neural network (CNN) based capacitance models (called CNN-Cap), which opens the third way for 3-D capacitance extraction to get accurate results with much less time cost than field solver. In this work, the techniques of neural architecture search (NAS) and data augmentation are proposed to train better CNN models for 3-D capacitance extraction.  Experimental results on datasets from different designs show that the obtained NAS-Cap models achieve remarkably higher accuracy  than CNN-Cap, while consuming less runtime for inference and space for model storage. Meanwhile, the  transferability of the NAS is validated, as the  once searched architecture brought similar
error reduction on coupling/total capacitance for the test cases from different design and/or process technology.
\end{abstract}



\maketitle

\section{Introduction}
Accurately modeling the interconnect parasitics (including resistance and capacitance) becomes more crucial for guaranteeing the performance of integrated circuits (ICs) \cite{choudhury1995automatic,lavagno2006eda,yu2021advancements}. 
As billions of transistors are placed within a chip, it is very challenging to perform full-chip capacitance extraction which computes all capacitance couplings among tens of billions  interconnect wires. An existing solution of this is the pattern matching approach widely used in commercial RCX tools.
It divides a large interconnect wire structure into small subregions or substructures, and then computes the capacitances of each substructure with pre-built empirical formulas or look-up tables of capacitance. The substructures sharing same geometry topology correspond to a pattern.
For a given process technology, a pattern library is pre-characterized by enumerating millions of sample structures and solving the capacitances for them with accurate field solver. Then, the empirical formulas or look-up tables can be obtained for the pattern structures, so that the capacitances of a substructure can be quickly computed at the time of full-chip extraction. 
However, the pattern matching  approach loses its accuracy due to limited coverage of interconnect typologies in real design or the error of empirical formulas, especially under the advanced process technologies. 

Another approach of capacitance extraction is based on field solver \cite{nabors1991fastcap,le1992stochastic,yu2013rwcap,wang2005improved,yu2014advanced,yu2016utilizing,yang2020floating,huang2024floating,yu2022advanced,yu2004pre,yu2012effi}, which has the highest accuracy. However, due to excessive computational cost, the field-solver based approach can only handle small structures and usually runs several orders of magnitudes slower than using the technique in the pattern matching approach.
In recent years, the machine learning or deep learning opens the third way for capacitance extraction. 
In \cite{kasai2019neural}, a neural network based method is presented to compute the capacitances of several structure patterns in three-dimensional (3-D) ICs.
Nevertheless, it only considers single-dielectric structures with simple multilayer perception (MLP) neural networks, and the demonstrated error on total capacitance can be larger than $10\%$ \cite{kasai2019neural}. 
Another approach based on MLP neural network was proposed in \cite{abouelyazid2022fast} for capacitance extraction of middle-end-of-line (MEOL) structures and interconnects with systematic process variations. It exhibits good accuracy for computing the capacitances in the regular structure of MEOL pattern.
Instead of directly computing capacitances, an MLP neural network based approach was proposed to improve the pattern matching based capacitance extraction through automatic pattern classification and capacitance formula building \cite{li2020layout}. A convolutional neural network (CNN) based approach (called CNN-Cap) for building models for computing capacitances of two-dimensional (2-D) pattern structures was proposed in \cite{yang2021cnn}. It employs a novel
grid-based data representation and leverages the CNN's ability of capturing spatial information to deliver more accurate models than using MLP neural network. Recently, the CNN-Cap was extended to compute the capacitances of 3-D structures using the ResNet architecture \cite{he2016deep} and facilitate the usage in practical full-net/full-chip extraction through the consideration of core-region of extraction window. The 3-D CNN-Cap exhibits good accuracy on predicting total capacitance of 3-D structures, i.e. with less than 5\% error in about 99\% probability, and runs 191X faster than the fast random walk based capacitance solver \cite{yu2013rwcap,yang2020floating}. 
In \cite{abouelyazid2022accuracy}, the DNN based approach for 3-D capacitance extraction was also proposed which is based on
a similar data representation to that in \cite{yang2021cnn,yang2023cnn}, but still employs the MLP neural network. Another model based on graph neural network (named GNN-Cap) was proposed for capacitance extraction very recently \cite{liu2024gnn}, which beats the RCX tool StarRC in terms of runtime and accuracy for extracting whole-net capacitances. However, the accuracy of GNN-Cap is much worse than CNN-Cap (see Fig. 17 and Table V in \cite{liu2024gnn}), and it seems GNN-Cap is suitable for estimating the whole-net capacitances, instead of the distributed capacitances.

Although CNN-Cap exhibits very good accuracy on 2-D structures, its
performance on 3-D structures is  not good enough, especially on coupling capacitances. Specifically, for a set of 3-D interconnect structures (with three metal layers), there are 4.1\% of coupling capacitances predicted by CNN-Cap exhibiting a relative error larger than 10\% \cite{yang2023cnn}. A significant limitation of CNN-Cap is that it only leverages the basic ResNet~\cite{he2016deep} architecture, without delving into the exploration of more potent neural architectures. Neural architectures assume a pivotal role in the realm of deep learning, especially considering that the selection of an effective neural architecture frequently hinges on the specific nature of the task at hand, such as YOLO~\cite{redmon2016you} in object detection. 
So in this work, we aim to obtain more accurate CNN models for 3-D capacitance extraction with the help of neural architecture search (NAS). NAS offers the capability to autonomously search an architecture that outperforms manually crafted counterparts, and it has witnessed rapid adoption across various domains, including image classification~\cite{chen2020drnas}, object detection~\cite{chen2019detnas}, and image segmentation~\cite{liu2019auto}. However, it has not been leveraged in the field of capacitance extraction.  

The major contributions of this work are as follows.

\begin{itemize}
    \item With the neural architecture search approach, we find a better CNN architecture than ResNet for the capacitance extraction of 3-D interconnects. It consists of normal cells and reduction cells with irregular operation graphs. 
    \item An approach of data augmentation for training the model for 3-D capacitance extraction is proposed. It inherits the physical nature of capacitance extraction and increases the training data by 8X with negligible extra training cost of time and storage.
    \item Combining the above techniques, we train the model named NAS-Cap as a successor of CNN-Cap.     
    Experimental results on 3-D interconnect structures have validated the effectiveness of the proposed techniques and demonstrated the advantages of the NAS-Cap models. For the dataset in \cite{yang2023cnn}, NAS-Cap produces the results with average error reduced by 1.8X and 1.5X on coupling capacitance and total capacitance respectively, compared to CNN-Cap. NAS-Cap also exhibits good transferability, as the once searched architecture brought similar error reduction on coupling/total capacitance for the test cases from different design and/or process technology. Meanwhile, the ratios of coupling capacitance with error larger than 10\% and total capacitance with error larger than 5\% are both remarkably reduced, thanks to the proposed NAS-Cap techniques. And, NAS-Cap model has 2.3X smaller storage size and 1.8X faster inference speed than CNN-Cap.
\end{itemize}

The rest of this article is organized as follows. The background of CNN-Cap model for 3-D capacitance extraction and the neural architecture search are introduced in Section 2.
In Section 3, we propose the techniques of NAS-Cap which improves CNN-Cap with neural architecture search and data augmentation. Then, numerical results are presented in Section 4. Finally, we draw the conclusion. Some preliminary results of this article were presented in \cite{li2024training}, in form of two-page extended abstract. We extend it with the details of neural architecture search, the training approach with data augmentation, and the experimental results showing the transferability of the searched CNN architecture.

\section{Background}
In this section, we first review the CNN-Cap method for 2-D and 3-D capacitance extraction. Then, the background of neural architecture search is briefly introduced.


\subsection{CNN-Cap Model for Capacitance Extraction}
CNN-Cap~\cite{yang2021cnn,yang2023cnn} is a convolutional neural network-based method for interconnect capacitance extraction. It was firstly proposed to extract the capacitances for  2-D pattern structure of cross-section view. Each structure includes at most three metal layers, and one master conductor and  $n_e$ environment conductors. One total capacitance and $n_e$ coupling capacitances of the master conductor are extracted. 
A grid-based data representation for the structure was proposed, which encodes the structure to several $L$-dimensional vectors, each for a metal layer. $L$ is a pre-defined value indicating the granularity of the grid division.
Specifically, a density vector $d \in \mathbb{R}^{3L}$ is first calculated, where the element $d_i$ indicates the ratio of conductor occupying a grid cell of the $\lceil i/L\rceil$-th metal layer. The encoded vector $x$ is first initialized to the density vector $d$. Then, if the grid cell for $d_i$ includes the master conductor, there is $x_i=d_i+1$. The resulting vector is used as an input to the sub-problem of extracting the total capacitance. For extracting the coupling capacitance related to one environment conductor, the environment conductor is similarly indicated in the encoding vector. For a 2-D structure used in training, $n_e+1$ vectors are generated as training data, along with one target value of total capacitance and $n_e$ target values of coupling capaictances.

\begin{figure}[h]
 \setlength{\abovecaptionskip}{1pt}
  \centering
    \includegraphics[width=3.54in]{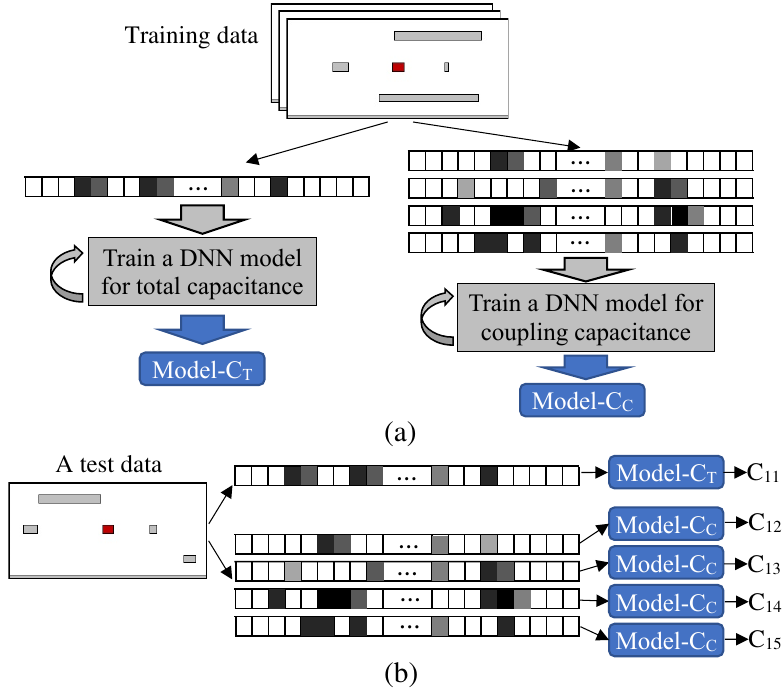}
  \caption{The process of building CNN-Cap models and using them for predicting 2-D capacitances. (a) The training stage. (b) The prediction stage  \cite{yang2023cnn}.
 \label{fig:approach}}
\end{figure}

After encoding the structure information, CNN-Cap utilizes a DNN model similar to ResNet to perdict the capacitance. Two models for the total capacitance and the coupling capacitance (called Model-C$_{\mathrm{T}}$ and Model-C$_{\mathrm{C}}$) are trained respectively. The target values of capacitance can be obtained from field solver, like Raphael \cite{raphaelv}. The process of building CNN-Cap models and using them to predict capacitances is illustrated in Fig.~\ref{fig:approach}.


\begin{figure}[!b]
\setlength{\abovecaptionskip}{1pt}
  \centering
    \subfigure[]
    {\includegraphics[width=1.7in]{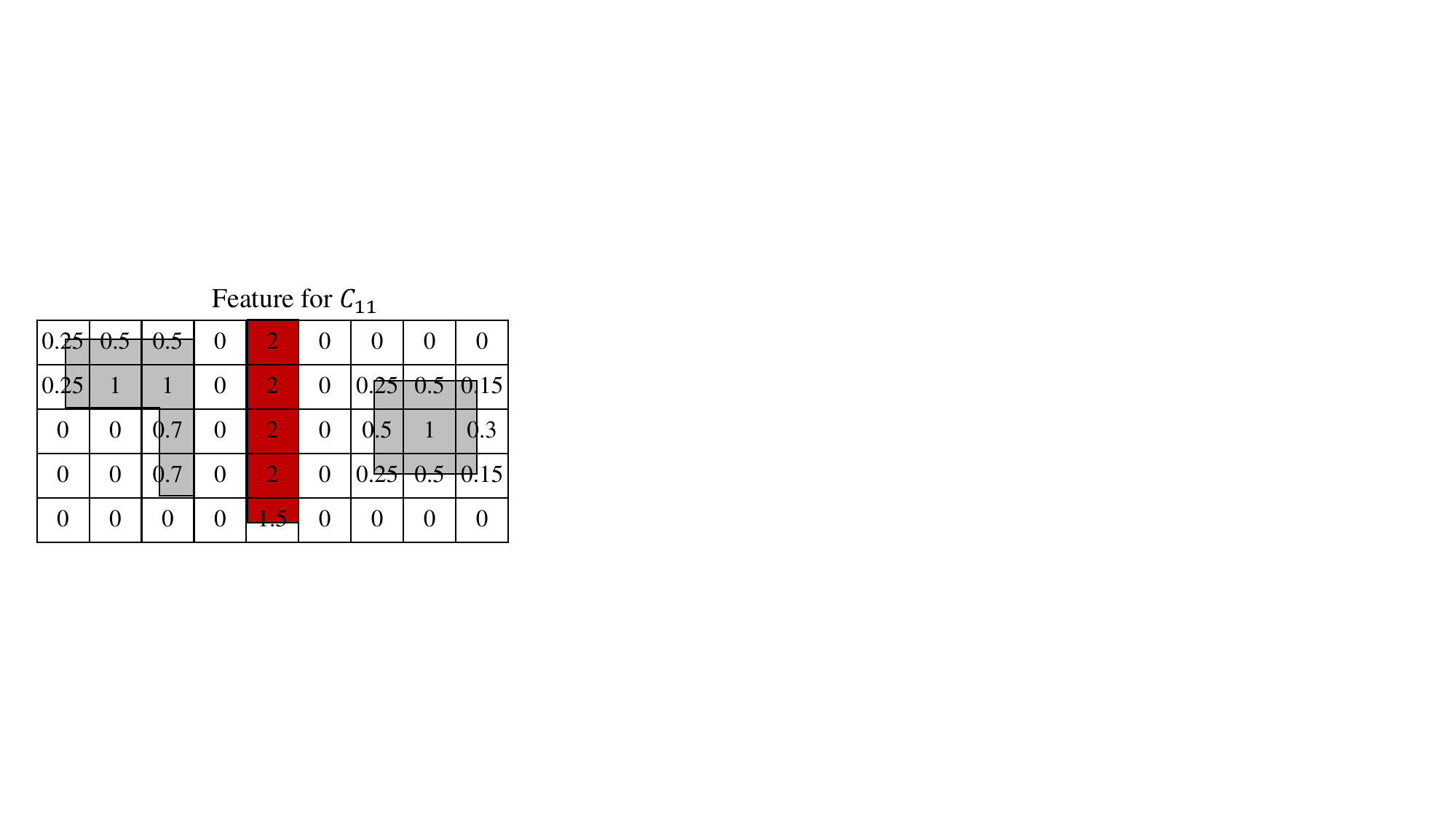}}\label{fig:exp0}  \hspace{0mm}
    \subfigure[]
    {\includegraphics[width=1.7in]{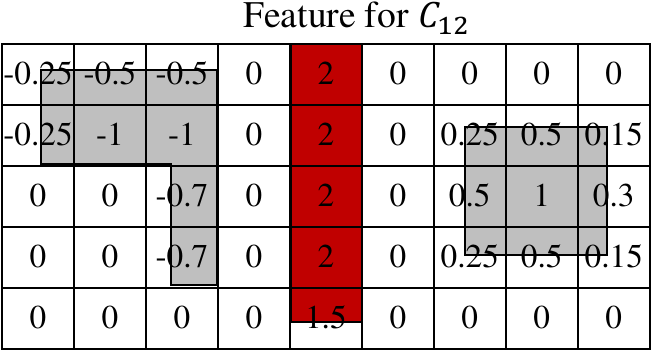}}\label{fig:exp1}  
  \caption{An example of a metal layer's data representation for 3-D interconnect structure. The case includes three conductors where the master conductor is in the middle. (a) The data representation for calculating the total capacitance $C_{11}$. (b) The data representation for calculating the coupling capacitance $C_{12}$ (Conductor 2 is on the left) \cite{yang2023cnn}.}
    \label{fig:3dfeature}
\end{figure}

The CNN-Cap was recently extended for 3-D capacitance extraction \cite{yang2023cnn}. The difference to 2-D problem is that a data becomes a 3-D window structure including interconnect wires on three metal layers (master conductor is in the middle layer). The data representation is similar. For each metal layer the data can be regarded as an $L\times  L$ 2-D grid, and the value for each grid cell is generated in same manner as in the 2-D CNN-Cap. An example is illustrated in Fig.~\ref{fig:3dfeature}, where for simplicity only one metal layer is shown. And, the DNN model used for 3-D problem is the basic ResNet \cite{he2016deep}. Experiments in \cite{yang2023cnn} show that it has good accuracy on predicting total capacitance, but the accuracy on coupling capacitance is not so good. For example, using ResNet-34 model, there are 4.1\% of testing data for which the error on coupling capacitance is larger than 10\%, and the  average relative error on coupling capacitance is 3.1\% \cite{yang2023cnn}. The CNN-Cap models and data have been shared on \cite{cnncapdata}.

\subsection{Neural Architecture Search}
The goal of neural architecture search (NAS) is to automatically design the neural architecture that achieves high performance. One of the pioneering works in the field of NAS is NAS-RL~\cite{zoph2017neural}, which employed reinforcement learning to search a model that can attain near state-of-the-art accuracy in the task of image classification. However, the method required a substantial amount of computational resources (i.e., 22,400 GPU days) on a small CIFAR-10 dataset~\cite{ren2021comprehensive}. To mitigate this computational burden, several techniques have been proposed to accelerate the search process, such as NASNet~\cite{zoph2018learning} and DARTS~\cite{liu2019darts}. Inspired by the fact that human-designed networks (e.g., ResNet) are often constructed from a repetitive stacking of a basic module, NASNet only searches a basic module (called cell), rather than directly searches a neural architecture. In DARTS, NAS is reformulated from a combinatorial optimization problem to a continuous differentiable optimization problem, which can be solved by gradient descent techniques. In this work, we leverage DrNAS~\cite{chen2020drnas}, which is an improved version of DARTS.

Apart from image classification, NAS has been employed in various other tasks, such as object detection~\cite{chen2019detnas} and image segmentation~\cite{liu2019auto}. However, to the best of our knowledge, no work has attempted to leverage NAS to search a neural architecture for capacitance extraction.

\section{Methodology}

In this section, we first propose the idea of using neural architecture search (NAS) to find a better CNN model for 3-D capacitance extraction. Then, we present the NAS inspired network model for capacitance extraction. Finally, we present the training skill based on a data augmentation approach for the obtained model.
\subsection{The Idea and Architecture Settings}
\begin{figure}[h]
    \centering
 \setlength{\abovecaptionskip}{1pt}
    \includegraphics[width=0.4\textwidth]{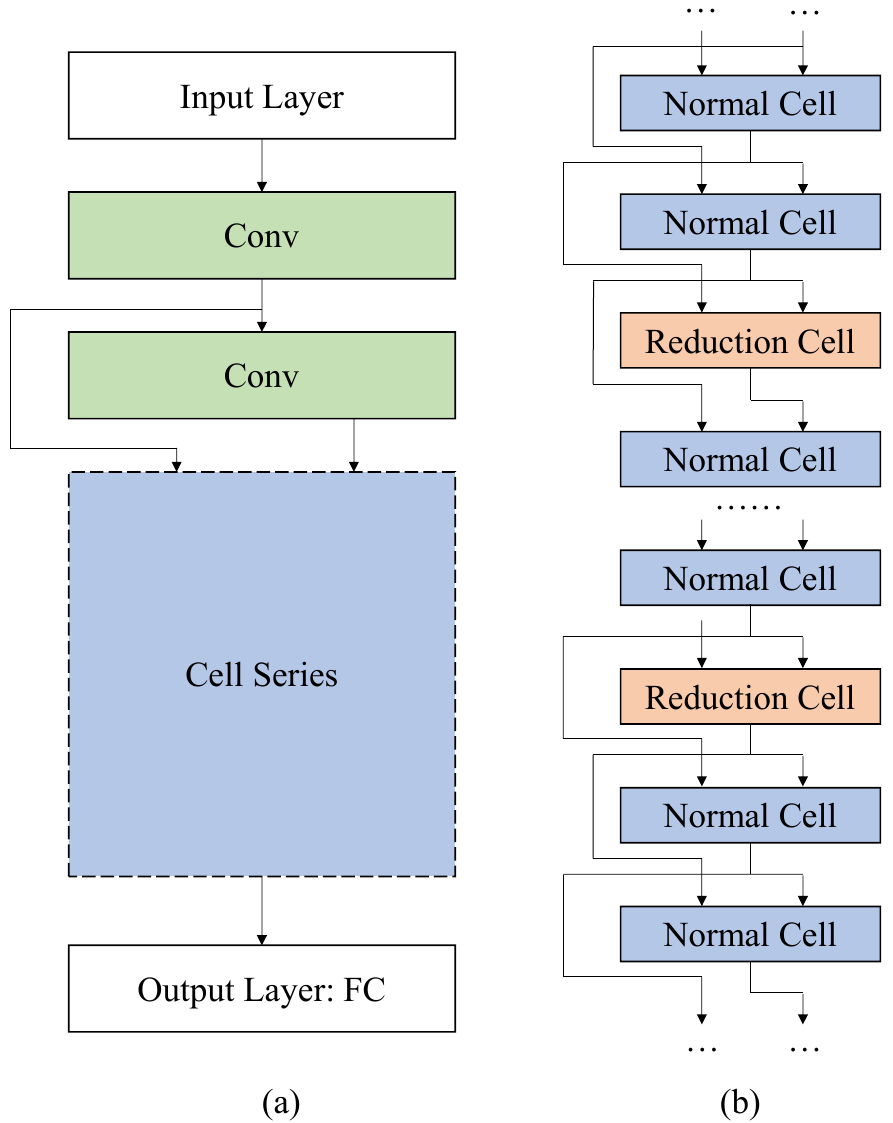}
    \caption{The architecture of the proposed model. (a) Overview of the whole network model, (b) Connection among cells in the cell series.}
    \label{fig:network}
\end{figure}

ResNet was originally designed for image classification. In order to find a more suitable network for capacitance extraction, we use the NAS method to automatically search the network structure. The first step for this to define a suitable search space, which is large enough but does not lead to huge computational resource for searching. Following DrNAS~\cite{chen2020drnas}, we use the cell-based search space. 
The considered network architecture is shown in Fig.~\ref{fig:network}(a), which consists of an input layer, 2 convolutional layers, a series of cells and an output layer. The connection among cells in the cell series
is shown in Fig.~\ref{fig:network}(b). As in ResNet and many other convolutional network, feature map shrinks its size and increases the number of channels while passing through some layers, we employ two kinds of cells: normal cell and reduction cell in the cell series. The reduction cells are the cells that halve the size of input data and double the number of channels, while normal cell keeps the shape of the input data. 
Each cell receives input data from the previous two cells, and its output data are fed to two subsequent cells. At the end of the cell series, the output data of the last cell are fed to the output layer. 
\begin{figure}[h]
  \setlength{\abovecaptionskip}{1pt}
   \centering
    \includegraphics[width=0.2\textwidth]{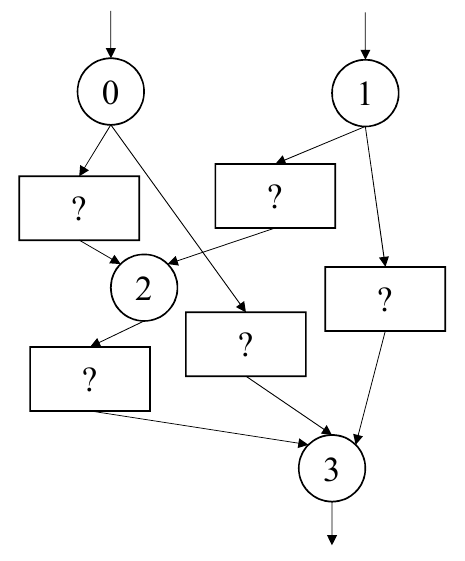}
    \caption{The directed acyclic graph (DAG) representing a cell, where each block with ``?'' represents an operation.}
    \label{fig:dag}
\end{figure}
Either normal cell or reduction cell can be represented as a directed acyclic graph (DAG) consisting of $N_d$ nodes, as is shown in Fig.~\ref{fig:dag}. Each node $x^{(i)}$ is an intermediate representation and each directed edge $(i, j)$ is associated with an operation $o^{(i,j)}$ that transforms $x^{(i)}$. The operations $o^{(i,j)}$ are selected from a candidate set $O$. 
Each intermediate node is computed as follows:
\begin{equation}
     x^{(j)} = \sum_{i<j}o^{(i,j)}(x^{(i)}) ~.
\end{equation}

It should be pointed out, that the architecture of ResNet can also be described as above. In ResNet, a residual block is equivalent to a cell. It consists of 2 stacked convultional layers, and a skip connecting adding up the input and the output of the convultional layer, forming the final output. An extra pooling layer is inserted for reduction cell.

In order to use gradient-based optimization method for searching the operations, the continuous relaxation is applied. An edge is now associated with a weighted sum of operations
\begin{equation}
     \hat{o}^{(i,j)}(x) = \sum_{o\in O}\theta^{(i,j)}_o o(x)  ~,
\end{equation}
where $\theta^{(i,j)}_o$ is the probability weight of each operation $o$, which satisfies the constraint $\sum_{o\in O}\theta^{(i,j)}_o =1$.
With continuous relaxation technique, the search process can be defined as a bi-level optimization problem:
\begin{equation}
\begin{aligned}
    &\min_\theta \mathcal{L}(w^*,\theta), \\
    \text{ s.t. } &w^*=\arg \min_w \mathcal{L}(w,\theta),
\end{aligned}
\end{equation}
where $\mathcal{L}(w^*,\theta)$ is the loss function, $\theta$ denotes the mixing weight of operations, and
$w$ denotes the parameters of the operation function. 
Notices that $\theta$ includes in $\theta^{(i,j)}_o$'s in (2). So, while solving (3) the following constraint should be enforced:
\begin{equation}
    \text{ s.t. }   \sum_{o\in O}\theta_o^{(i,j)}=1, ~\forall (i,j), i<j,
\end{equation}
In the process of solving (3), $w$ and $\theta$ are optimized by turns, fixing one while optimizing the other. 
Directly optimizing $\theta$ with gradient descent will probably break the constraint.
To maintain the constraint of $\theta$ during optimization, we can substitute $\theta$ with $\theta=\text{Softmax}(\alpha)$. Softmax()  takes any vector as input and sum of the result elements equals 1. So, we can optimize $\alpha$ instead and will not break the constraint.

In this work, the operation set $O$ contains common building blocks of convolutional networks, such as convolutional layers of different size, max pooling and average pooling. Besides, there are two special operations, identity operation and zero operation. The zero operation indicates that there is no connection between two node. As for the convolutional layers, a RELU layer is inserted before each of them.
Specifically, the operation set contains $3 \times 3$, $5 \times 5$ and $7 \times 7$ normal convolutions, $3 \times 3$ and $5 \times 5$ separable convolutions, $3 \times 3$ and $5 \times 5$ dilated separable convolutions, $3 \times 3$ max pooling, $3 \times 3$ average pooling, identity, and zero operations (11 in total). And, for each operation there are two versions, one with batch normalization (BN) layer and the other without BN layer.






\subsection{NAS-Cap Model for Capacitance Extraction}
The work flow of using NAS to obtain a better model for capacitance extraction is as follow. At first, the operation parameters $w$ and the mixing weight $\theta$ are optimized simultaneously. During searching process (i.e. solving (3) as explained in last subsection), we prune the operation set progressively, keeping the operations with high mixing weight, in order to speed up the searching process. After sufficient steps, we pick the operation $o$ with the highest mixing weight $\theta_{o}$ to form the result cell. Then, we build the network with the cells and retrain the operation parameters $w$.

\begin{figure}[h]
 \setlength{\abovecaptionskip}{1pt}
    \centering
    \includegraphics[width=0.75\textwidth]{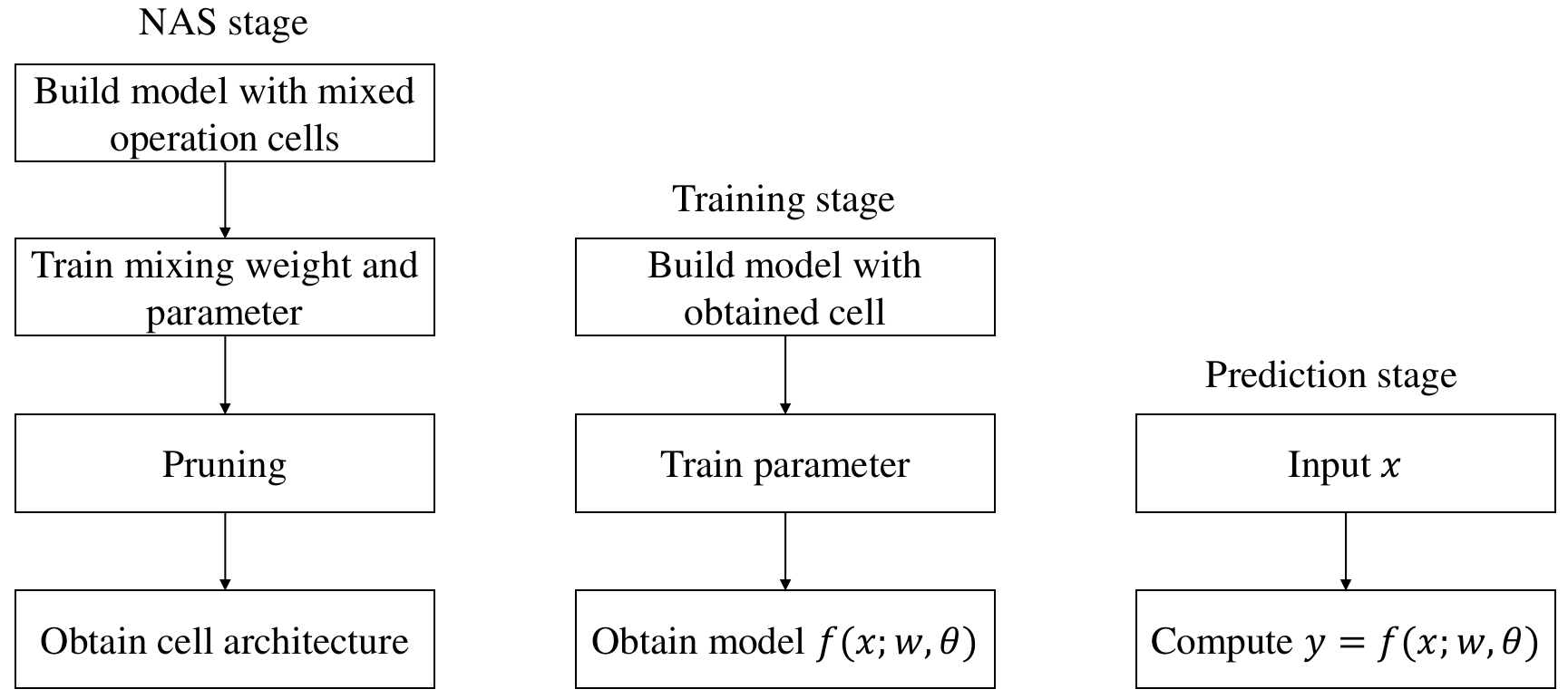}
    \caption{The workflow of the proposed deep-learning based method with neural architecture search.}
    \label{fig:stage}
\end{figure}

As for the loss function, we use mean square relative error (MSRE), whose expression is as follow:
\begin{equation}
    \mathcal{L}(w,\theta)=\frac{1}{N} \sum_{i=1}^N (1-\frac{f(\boldsymbol{x}^{(i)};w,\theta)}{\boldsymbol{y}^{(i)}})^2 ~,
\end{equation}
where $N$ is the number of training data, $f(\cdot;w,\theta)$ is the neural network, $\boldsymbol{x}^{(i)}$ indicates the $i$-th input data, and $\boldsymbol{y}^{(i)}$ is the corresponding label. This MSRE function includes the relative error, which can attain same accuracy on capacitance of different orders of magnitude.
Once the NAS is completed, the architecture, i.e. parameters $\theta$, is fixed. Then, the model represented by $w$ is retained with the loss function (4) with fixed $\theta$. Finally, the model for capacitance extraction is obtained, which can be used for prediction. The whole workflow of the deep-learning based method with neural architecture search is shown in Fig. \ref{fig:stage}.

\begin{figure}[h]
 \setlength{\abovecaptionskip}{1pt}
  \centering
    \includegraphics[width= 0.54\textwidth]{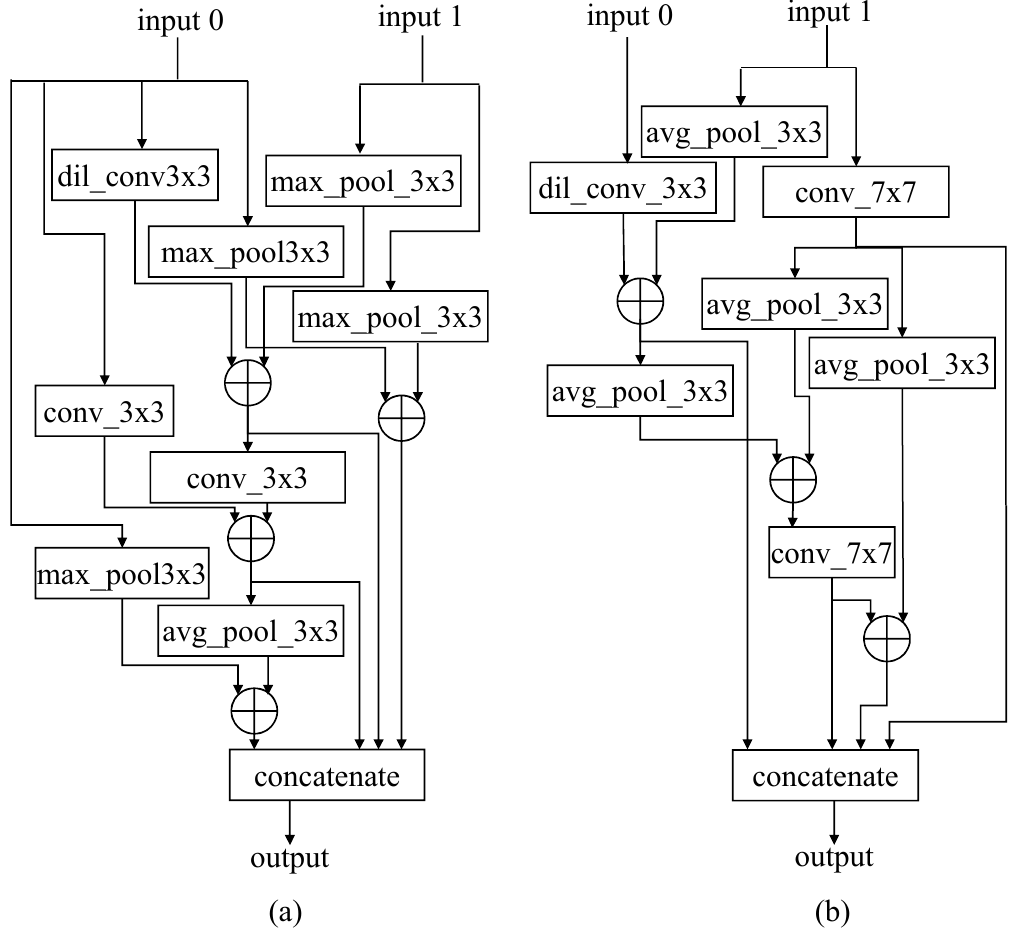}
  \caption{Architecture of a cell in NAS-Cap. (a) Normal cell. (b) Reduction cell.}
      \label{fig:cell}
\end{figure}

Directly doing NAS with the settings in last subsection, 
we found out that the resulted structure do not have BN layer in most cells. It is different from that the original ResNet or CNN-Cap. We further find out that, recent researches have shown that while BN performs well in computer vision, it performs poorly in some other tasks such as neural machine translation~\cite{shen2020powernorm}. There are some work attempting to explain this phenomenon empirically or theoretically. For example, in~\cite{wang2022understanding}, it was observed that the training loss and the validation loss have an inconsistent tendency during training when using the BN, which usually harms the performance.
This inspired us setting each operation in Fig. 4 without BN and rerunning the NAS. Our experiments show that the obtained model has better performance than the searched model under the setting of keeping all BN layers. 
The resulted network model has the normal cell and reduction cell shown in  Fig.~\ref{fig:cell}. They are more complicated than the blocks in ResNet. This resulted model is called NAS-Cap.

\subsection{Training  NAS-Cap Models with Data Augmentation}
\begin{figure}[h]
  \setlength{\abovecaptionskip}{1pt}
   \centering
    \includegraphics[width=0.42\textwidth]{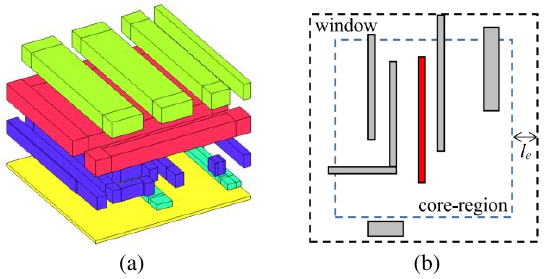}
    \caption{(a) A 3-D structure for capacitance extraction, and (b) A top view of the metal layer including the master conductor (from \cite{yang2023cnn}).}
    \label{fig:3dw1}
\end{figure}

Once the CNN architecture is obtained, we use it to train two NAS-Cap models for extracting total capacitance and coupling capacitance, respectively. 
While training NAS-Cap model, the input data corresponds to the 3-D window structure, such as that shown in Fig.~\ref{fig:3dw1}, with its capacitance values as labels. Once the master conductor is set, the 3-D window structure is converted to the grid-based data representations with the approach in Section 2.1. They correspond to the tasks of extracting one total capacitance and a few of coupling capacitances.
To obtain the labels, 3-D field solver is needed, causing large time consumption for collecting sufficient training data. Therefore, a data augmentation approach is favorable if it can improve the  model's performance without increasing the cost of data preparation.

\begin{figure}[h]
 \setlength{\abovecaptionskip}{1pt}
    \centering
    \includegraphics[width=0.5\textwidth]{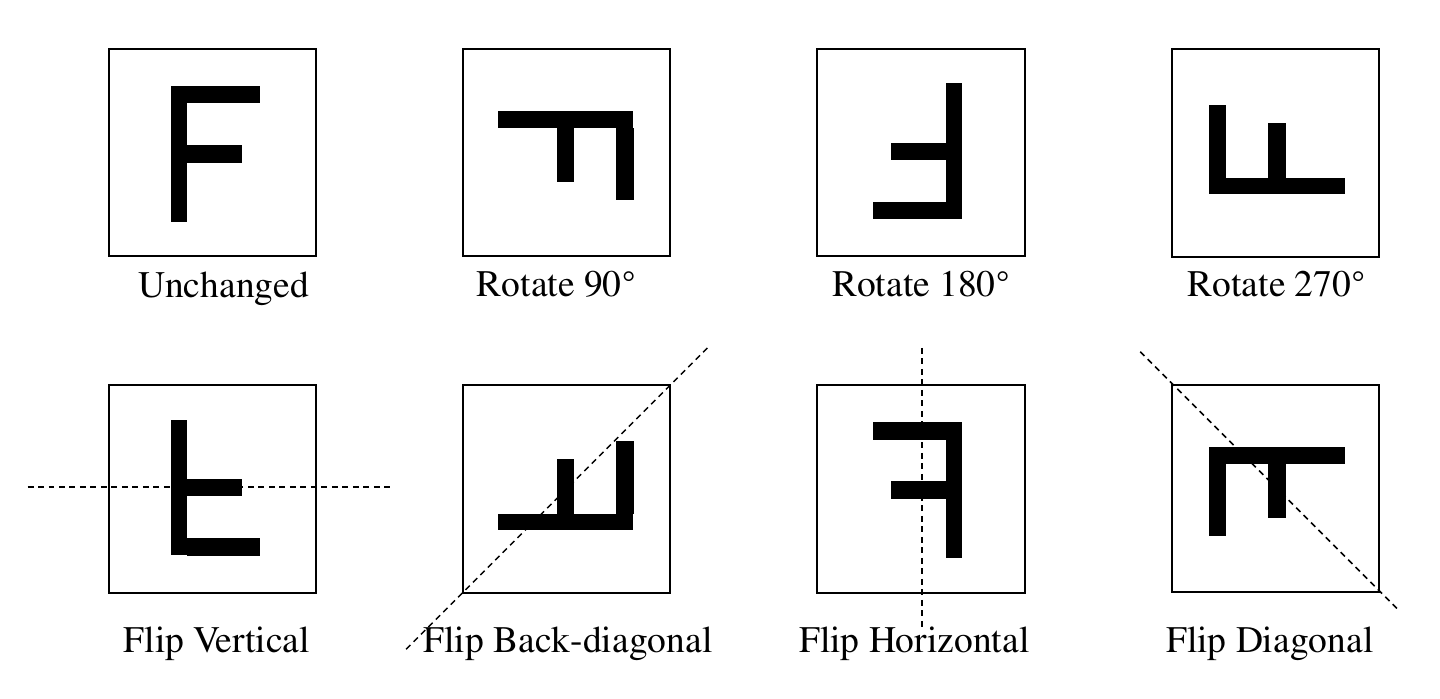}
    \caption{An original 3-D interconnect structure (in top view) for capacitance extraction, and its 7 equivalences after performing 7 geometry transformations.}
    \label{fig:data_aug}
\end{figure}

The physical nature of capapcitance extraction problem actually provides such a good data augmentation approach. As the sidewalls of the the structure are all set Neumann boundary condition. The structure's capacitances will not change after performing 7 transformations shown in Fig.~\ref{fig:data_aug}. 
And, the transformed structure usually corresponds to different grid-based data representations. 
This approach increases the training data by 8X with negligible effort. In order to reduce the storage cost for all the training data after augmentation, we further propose not to explicitly generate all the augmented data. Instead, during training the NAS-Cap model, when we fetch a data we take one of the eight equivalent forms (in Fig. 7) randomly and then feed it into the model. 
This improve the diversity of data, and does not make the training time longer. Experimental results show that this data augmentation approach helps to increase the accuracy of the trained model, while preserving the efficiency of time and storage.

\section{Experimental Results}

%
We carry out the experiments on three datasets from our industrial partner: the first is the dataset for 3-D extraction experiment 
 in~\cite{yang2023cnn,cnncapdata} and the other two are new datasets.
The dataset in~\cite{cnncapdata} includes  8685 3-D interconnect structures obtained from a real SRAM design with layout size $165\mu m \times 188\mu m $. Each structure is a $5\mu m \times 5\mu m $ window randomly cut from the layout. Setting the metal-layer combination (1, 2, 3) and different master conductors in core-region (see Fig.~\ref{fig:3dw1}), it further results in 8049 valid structures and  13579 data for extracting total capacitance and 72226 data for extracting coupling capacitance. Notice that the master conductor is on the 2nd metal-layer for this dataset.
The second dataset is derived from the same 8685 structures as Dataset 1, but the  metal-layer combination is changed to (2, 3, 4). The last dataset is from an analog circuit design with layout size  $749\mu m \times 1898\mu m$.
The detailed information of these datasets are listed in Table 1. For Dataset 3, the window size is $5\mu m \times 5\mu m $, and in order to control the data size we only randomly take in 1/5 of the data for coupling capacitances.   
\begin{table}[htbp]
\centering
 \setlength{\abovecaptionskip}{1pt}
\caption{The Datasets for Validating the Proposed NAS-Cap for 3-D Capacitance Extraction}\label{table-data}
\begin{tabular}{|c|c|c|c|c|c|}
\hline
Dataset & Layout Size & Layer Combination & \# Sample Structure & \# Total Cap. & \# Coupling Cap. \\
\hline
1 &$165\mu m \times 188\mu m $ & (1, 2, 3)  & 8049  & 13579 & 72226 \\
\hline
2 &$165\mu m \times 188\mu m $ & (2, 3, 4) & 1400 & 2475 & 13072 \\
\hline
3 & $749\mu m \times 1898\mu m $& (2, 3, 4) & 21566 & 53000 & 38371 \\
\hline
\end{tabular}
\end{table}


We compare the performance of proposed NAS-Cap and CNN-Cap \cite{yang2023cnn} based on these datasets. During NAS, the number of DAG nodes is set to 6, and the number of cells is 14 in the cell series. The learning rates for training the architecture parameters $\theta$ and network parameters $w$ are set to 0.006 and 0.0001, respectively. 
Batch size of both stages are 32. At the NAS stage, the number of training epochs is 90, and is evenly divided into 3 sub-stages. Between sub-stages, the size of the operation set is pruned from 11 to 7, then to 4. 

The NAS is performed only on Dataset 1. Then, the obtained NAS-Cap models are trained for each dataset. Each dataset is split into a training subset (with 90\% of the samples) and a testing subset (with 10\% of the samples).
At the model training stage, the number of epochs is 250, and the cosine annealing strategy is applied to learning rate in order to make the optimization process converge.


The DNN models are implemented with PyTorch, and all experiments are carried out on a Linux server with two Intel Xeon Silver 4214 CPUs at 2.2GHz and 4 Nvidia RTX4090 GPUs.

\subsection{Results on Dataset 1}

With the obtained NAS-Cap architecture (described by Fig. 3 and Fig. 6), the DNN model is trained for predicting the capacitances for Dataset 1. The prediction errors (for the testing subset) on coupling capacitance and total capacitance are shown in Table 2 and Table 3, respectively. 
The results of NAS-Cap models without and with the data augmentation (in Section 3.3) are given, and compared with those of CNN-Cap. 
Here, $Err_{avg}$ means the average of the relative error's absolute value, $Err_{max}$ means the maximum of the relative error's absolute value, and Ratio (Err>10\%) means the ratio of the number of coupling capacitances with error larger than 10\% to the number of all coupling capacitances predicted. 
From Table 2, we see that with the proposed NAS-Cap model most of coupling capacitances have error within 10\%; only for 0.74\% of testing data the error of coupling capacitance is larger than 10\%. And, the average error is just 1.7\%. Although the maximum error is similar to that of CNN-Cap, for most cases the error is large reduced to within 10\%. This is also revealed in Fig. 9(a). From the results  we can also easily see that the individual benefits of using NAS and the data augmentation approach.

\begin{table}[htbp]
\centering
 \setlength{\abovecaptionskip}{1pt}
\caption{Prediction Error on Coupling Capacitance for the Testing Subset of Dataset 1}\label{table-cp}
\begin{tabular}{|c|c|c|c|}
\hline
\multirow{2}*{Method}&\multicolumn{3}{|c|}{Error on Coupling Capacitance} \\
\cline{2-4} 
  & $Err_{avg}$& $Err_{max}$& Ratio (Err$>$10\%) \\
\hline
CNN-Cap \cite{yang2023cnn}& 3.1\%& 44\% & 4.1\% \\
\hline
NAS-Cap w/o DA & 2.4\% & 42\% & 2.2\% \\
\hline
NAS-Cap& 1.7\%	& 58\%	&	0.74\% \\
\hline
\end{tabular}
\end{table}
\begin{table}[htbp]
\centering
 \setlength{\abovecaptionskip}{1pt}
\caption{Prediction Error on Total Capacitance for the Testing Subset of Dataset 1}\label{table-tot}
\begin{tabular}{|c|c|c|c|}
\hline
\multirow{2}*{Method} &\multicolumn{3}{|c|}{Error on Total Capacitance} \\
\cline{2-4} 
 & $Err_{avg}$& $Err_{max}$& Ratio (Err$>$5\%) \\
\hline
CNN-Cap \cite{yang2023cnn}& 1.1\%& 19\% & 1.3\% \\
\hline
NAS-Cap w/o DA& 0.84\%& 22\% & 0.74\% \\
\hline
NAS-Cap & 0.74\%& 13\% & 0.30\% \\
\hline
\end{tabular}
\end{table}

From Table 3, we see that the proposed approaches also demonstrate the remarkably better accuracy than CNN-Cap. The ratio of total capacitance whose error is larger than 5\% is reduced from 1.3\% to just 0.3\%. 
The corresponding error distribution is shown in Fig. 9(b). 

\begin{figure}[]
 \setlength{\abovecaptionskip}{1pt}
  \centering
    \subfigure[]
    {\includegraphics[width=0.48\textwidth]{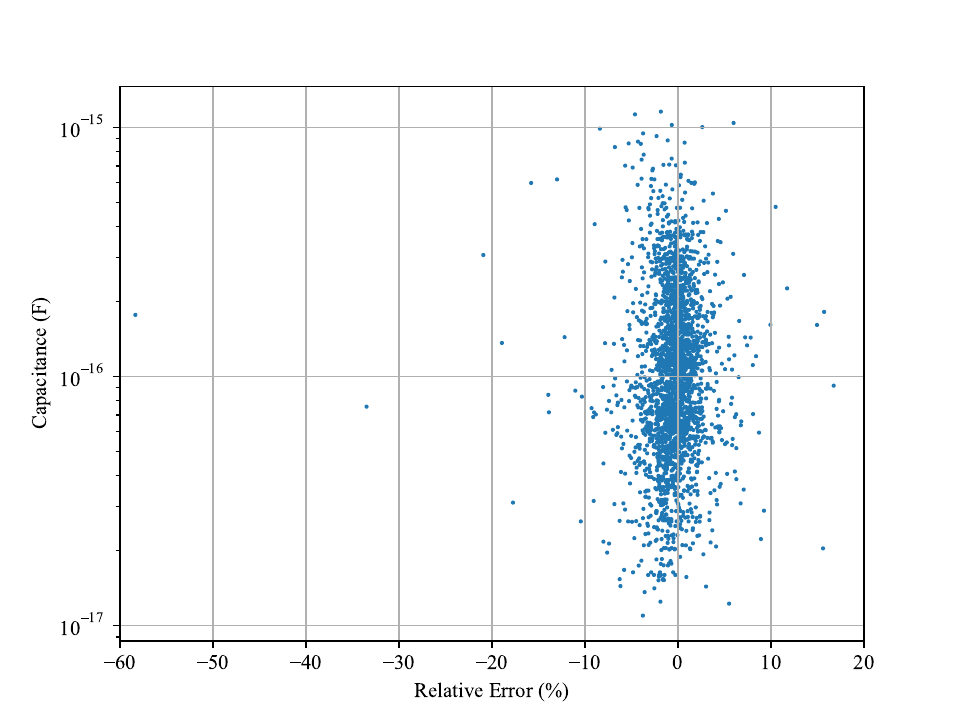}\label{fig:total}} 
    \subfigure[]
    {\includegraphics[width=0.48\textwidth]{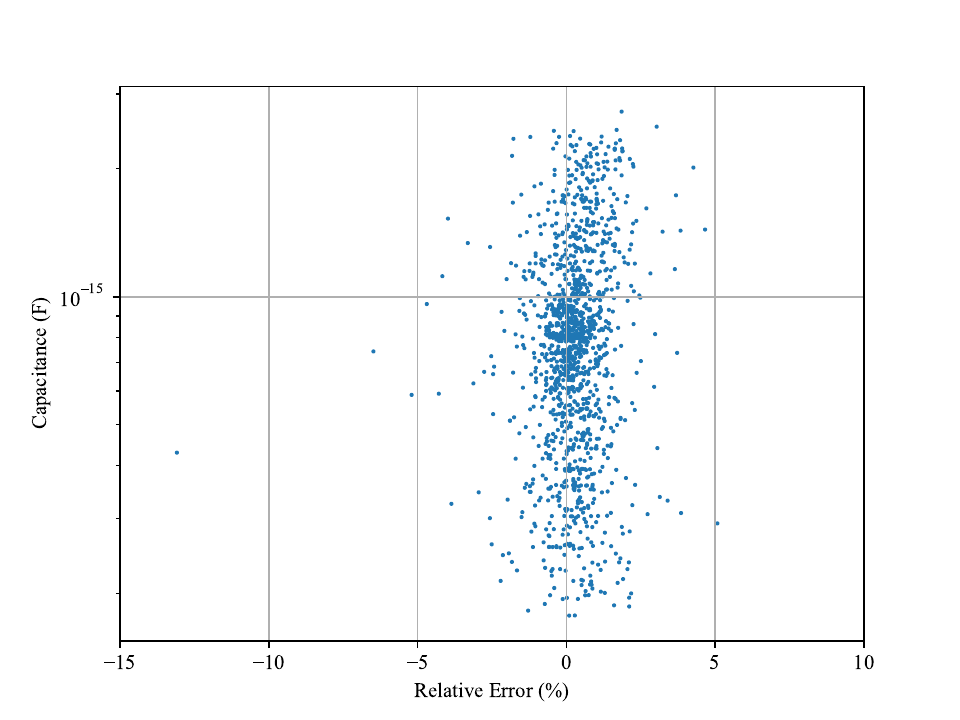}\label{fig:env}}  
  \caption{The capacitance result of NAS-Cap versus relative error for 3-D structures in Dataset 1. (a) Results of coupling capacitance. (b) Results of total capacitance.}
      \label{fig:accuracy}
\end{figure}

We now compare the inference time and model size of NAS-Cap model and CNN-Cap model. The results are listed in Table 4. From it we see that the inference time of NAS-Cap model is about half of that of CNN-Cap model. As for the model size, the NAS-Cap model has 9,906,697 parameters, occupying 39.6 MB storage, which is less than a half of the size of CNN-Cap model.

The training time for NAS-Cap model is no more than 4 hours. As for the architecture search, it costs about 16 hours, which can be amortized for different NAS-Cap models for different metal-layer combinations. As we will show later, the CNN architecture obtained from NAS has good tranferability, which performs well for structures from different design and even different process technology.
\begin{table}[htbp]
\centering
 \setlength{\abovecaptionskip}{1pt}
\caption{The Average Inference Time Per Case and Model Storage Size}\label{table-time}
\begin{tabular}{|c|c|c|}
\hline
Method& Time (ms)& Storage Size (MB) \\
\hline
CNN-Cap \cite{yang2023cnn}& 0.11& 89.7 \\
\hline
NAS-Cap & 0.06& 39.6 \\
\hline
\end{tabular}
\end{table}

\subsection{Results on Dataset 2 and Dataset 3}

In this subsection, we present the experimental results on Dataset 2 and Dataset 3, which show the transferability of the searched CNN architecture. Notice that Dataset 3 is constructed from a design under a different process technology.
On these two datasets, we train the NAS-Cap models respectively, and then evaluate their performance on the corresonding testing subsets.
The prediction errors on coupling capacitance and total capacitance are listed in Table 5 and Table 6, respectively. 
The capacitance results from NAS-Cap models and their relative errors are plotted in Fig. 10 and Fig. 11.
From them we can see that, compared to CNN-Cap models, the NAS-Cap models reduce the error on coupling capacitance by nearly 2X, and the ratio of coupling capacitances with error larger than 10\% by more than 2X. While for extracting total capacitance, the NAS-Cap is also better than CNN-Cap, though the advantage is not very large.
For Dataset 3, the error of NAS-Cap is relatively large, compared to other two datasets. Even though, the average error on coupling capacitance is reduced to just 3.5\%. 

\begin{table}[htbp]
\centering
 \setlength{\abovecaptionskip}{1pt}
\caption{Prediction Error on Coupling Capacitance for the Testing Subsets of Dataset 2 and 3}\label{table-cp}
\begin{tabular}{|c|c|c|c|c|}
\hline
& \multirow{2}*{Method} &\multicolumn{3}{|c|}{Error on Coupling Capacitance} \\
\cline{3-5} 
& & $Err_{avg}$& $Err_{max}$& Ratio (Err$>$10\%) \\
\hline
\multirow{2}*{Dataset 2} & CNN-Cap \cite{yang2023cnn}& 3.7\% & 75\%  &7.5\%\\
\cline{2-5} 
& NAS-Cap& 2.3\% & 21\% & 2.2\% \\
\hline
\multirow{2}*{Dataset 3} & CNN-Cap \cite{yang2023cnn}& 6.0\% & 160\%	& 16\% \\
\cline{2-5} 
& NAS-Cap & 3.5\%	& 288\%	&	7.1\% \\
\hline
\end{tabular}
\end{table}

\begin{table}[htbp]
\centering
 \setlength{\abovecaptionskip}{1pt}
\caption{Prediction Error on Total Capacitance for the Testing Subsets of Dataset 2 and 3}\label{table-tot}
\begin{tabular}{|c|c|c|c|c|}
\hline
& \multirow{2}*{Method} &\multicolumn{3}{|c|}{Error on Total Capacitance} \\
\cline{3-5} 
& & $Err_{avg}$& $Err_{max}$& Ratio (Err$>$5\%) \\
\hline
\multirow{2}*{Dataset 2} & CNN-Cap \cite{yang2023cnn}& 1.9\%	&11\% & 8.5\% \\
\cline{2-5} 
& NAS-Cap & 1.6\% & 8.2\% &8.0\% \\
\hline
\multirow{2}*{Dataset 3} & CNN-Cap \cite{yang2023cnn}& 1.3\%	& 27\% & 4.3\% \\
\cline{2-5} 
& NAS-Cap & 0.99\% & 30\% &2.2\% \\
\hline
\end{tabular}
\end{table}


\begin{figure}[h]
 \setlength{\abovecaptionskip}{1pt}
  \centering
    \subfigure[]
    {\includegraphics[width=0.48\textwidth]{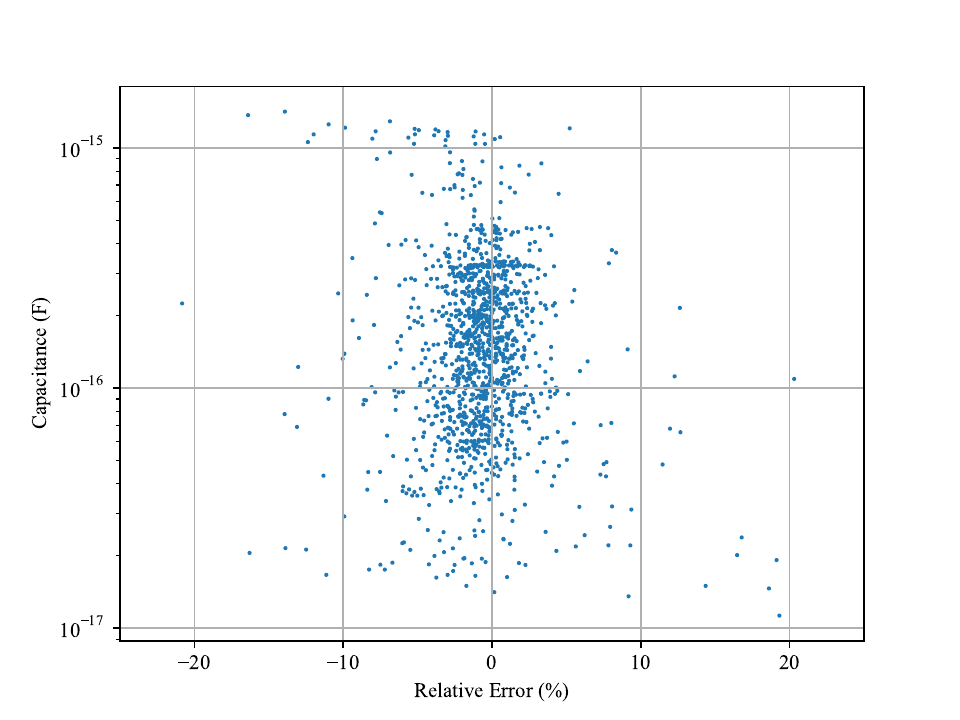}\label{fig:total}} 
    \subfigure[]
    {\includegraphics[width=0.48\textwidth]{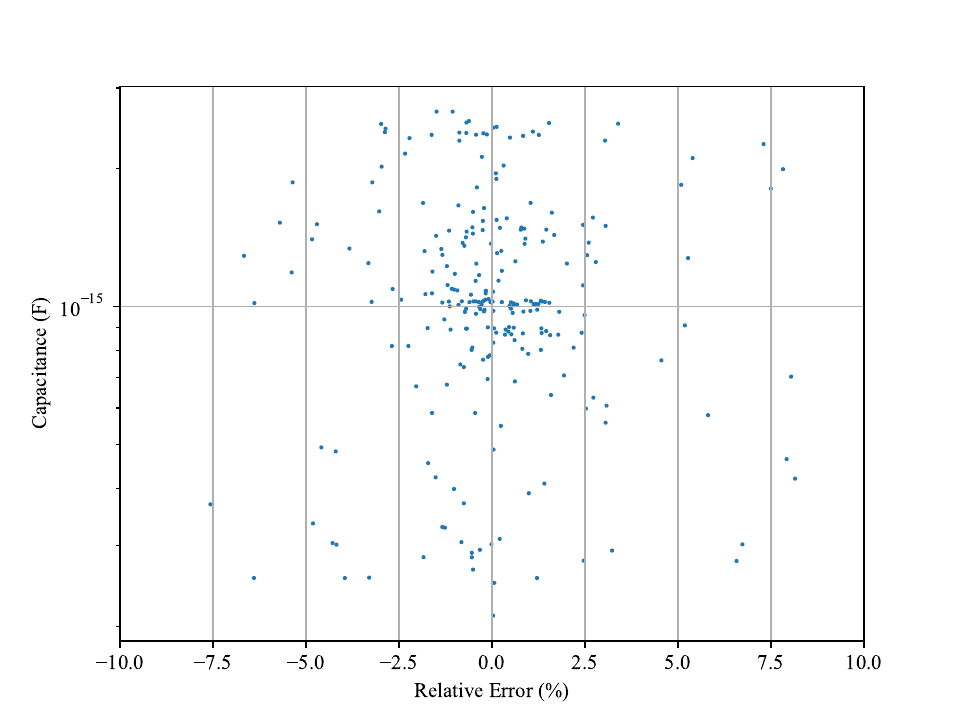}\label{fig:env}}  
  \caption{The capacitance result of NAS-Cap versus relative error for 3-D structures in Dataset 2. (a) Results of coupling capacitance. (b) Results of total capacitance.}
      \label{fig:accuracy}
\end{figure}

\begin{figure}[h]
 \setlength{\abovecaptionskip}{1pt}
  \centering
    \subfigure[]
    {\includegraphics[width=0.48\textwidth]{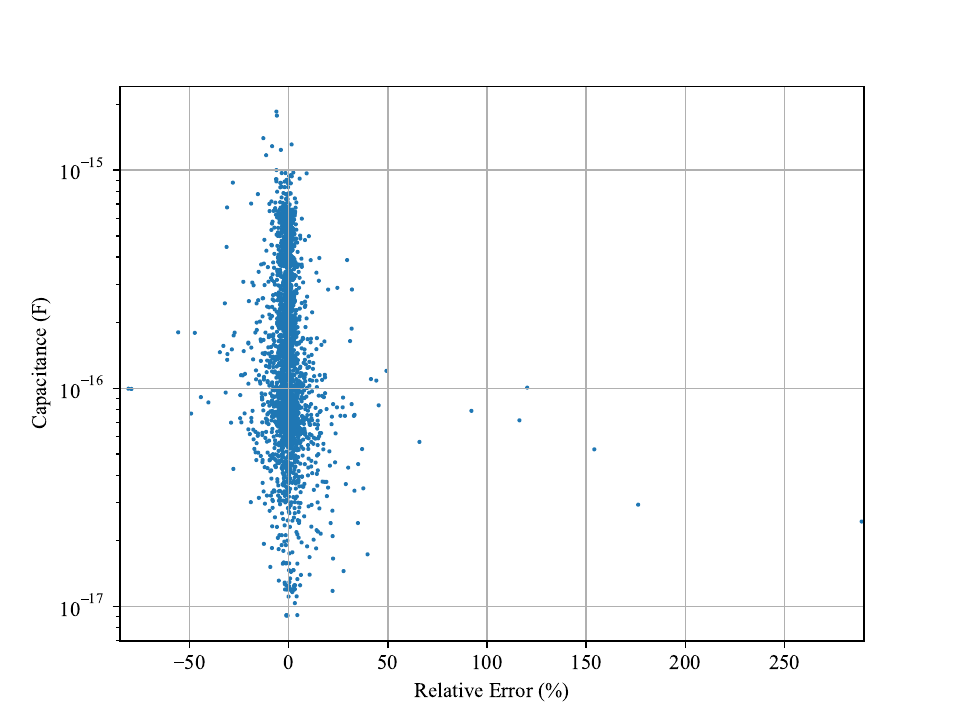}\label{fig:total}} 
    \subfigure[]
    {\includegraphics[width=0.48\textwidth]{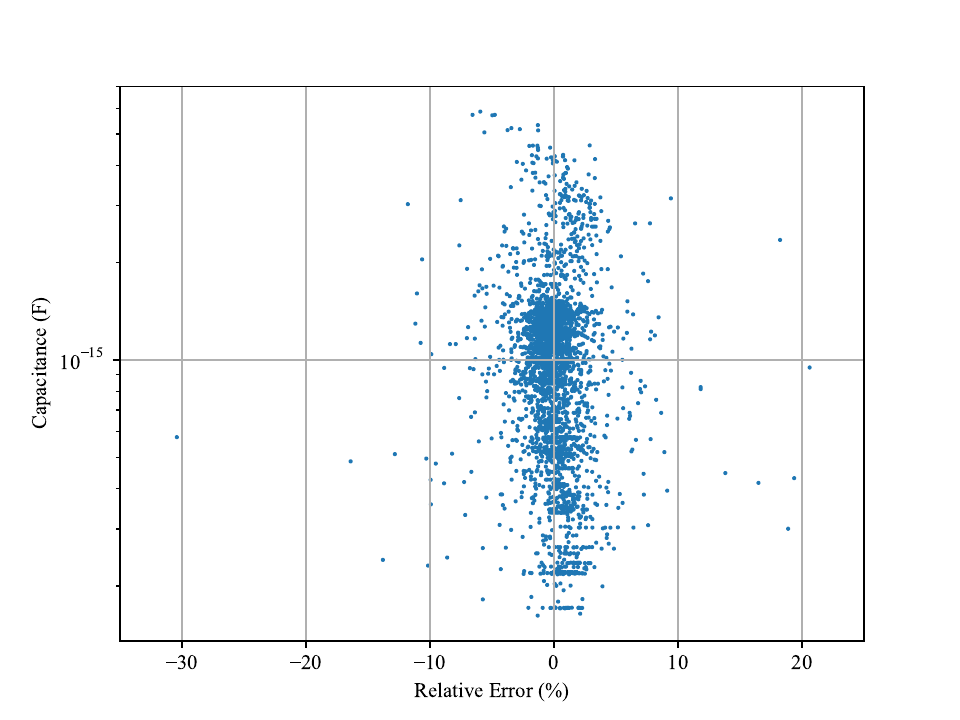}\label{fig:env}}  
  \caption{The capacitance result of NAS-Cap versus relative error for 3-D structures in Dataset 3. (a) Results of coupling capacitance. (b) Results of total capacitance.}
      \label{fig:accuracy}
\end{figure}

Regarding the model size and inference time, on these datasets NAS-Cap models show same advantage over CNN-Cap models as that shown in Table 4.

At the end of this section, we summarize the experimental results on the three datasets.
\begin{itemize}
    \item The proposed techniques (neural architectural search and data augmentation) both improve the CNN-Cap model \cite{yang2023cnn} in terms of accuracy of capacitance extraction. Especially on coupling capacitance, the error is reduced by 1.8X, 1.6X and 1.7X respectively on the three datasets. Meanwhile, the error distribution range is largely narrowed, as demonstrated by the last columns in Table 2 and 5.
    \item The proposed NAS-Cap models are also beneficial for extracting total capacitance. Compared to CNN-Cap models, the error on total capacitance is reduced by 1.5X, 1.2X and 1.3X respectively on the three datasets. The error distribution range is also narrowed, as demonstrated by the last columns in Table 3 and 6.
    \item The NAS-Cap model has less than half size of storage compared to CNN-Cap model, and runs about 2X faster than CNN-Cap for inferring capacitances.
    \item The searched DNN architecture from NAS has good transferablity, which means it always shows remarkable advantages over the CNN-Cap architecture for extracting capacitances from different designs and different process technologies. This validates the significance of performing NAS.
\end{itemize}

\section{Conclusions}
In this work, we present an approach of discovering better CNN models for capacitance extraction of 3-D interconnect structures with the help of neural architecture search and an approach of data augmentation. Experiments show that the obtained NAS-Cap model largely improves the accuracy of predicted capacitances and the efficiency of inference. The transferability of the NAS is also validated.
The proposed approach is expected to better revamp the accuracy issue of the pattern matching based approach for full-chip extraction. And, how to collaborate the NAS-Cap with the pattern-matching based full-chip extraction could be investigated in the future. 



\bibliographystyle{ACM-Reference-Format}
\bibliography{refs}










\end{document}